\documentclass[12pt]{article}

\title{Neutron-Mirror-Neutron Oscillations in a Trap\\}
\author{B.Kerbikov\thanks{E-mail:borisk@itep.ru} and O.Lychkovskiy\thanks{E-mail:lychkovskiy@itep.ru}\\
 State Research
Center\\Institute for Theoretical and Experimental Physics, \\
Moscow, Russia}
 \date{}

\newcommand{\beq}{\begin{eqnarray}}
 \newcommand{\eeq}{\end{eqnarray}}
  \newcommand{\be}{\begin{equation}}
\newcommand{\ee}{\end{equation}}
\def\la{\mathrel{\mathpalette\fun
<}} 
\def\fun#1#2{\lower3.6pt\vbox{\baselineskip0pt\lineskip.9pt
\ialign{$\mathsurround=0pt#1\hfil ##\hfil$\crcr#2\crcr\sim\crcr}}}

\newcommand{\lan}{\langle}
\newcommand{\ran}{\rangle}

\begin{document}

\maketitle
\begin{abstract}
We calculate the rate of neutron-mirror-neutron oscillations for
ultracold neutrons trapped in a storage vessel. Recent
experimental bounds on the oscillation time are discussed.
\end{abstract}
\section{Introduction}

During the last couple of years we are witnessing the revival of
the interest to the "mirror particles",  "mirror matter" and
"mirror world". The idea of the existence of the hypothetical
hidden sector to compensate mirror asymmetry was first explicitly
formulated in \cite{1}. The subject has a rich history -- see the
review paper \cite{2}. The present wave of interest to mirror
particles has been to a great extent initiated by  the quest for
neutron-mirror-neutron oscillations ($n-n')$. It was conjectured
that $n-n'$ oscillations may play an important role in the
propagation of ultra high energy cosmic rays and that the
oscillation time $\tau_{osc}$ may be as small  as $\tau_{osc} \sim
1 s$ \cite{3}.  Implications of mirror particles for cosmology and
astrophysics were discussed in a  number of papers e.g.,
\cite{4a}. Last year the first experimental data on $n-n'$
transitions were published with the results $\tau_{osc}\geq 103s$
\cite{4} and $ \tau_{osc}\geq 414s$ \cite{5}. Possible laboratory
experiments to search for $n-n'$ oscillations were discussed in
\cite{6}.

Experimental results \cite{4,5} were obtained using the ultracold
neutrons (UCN), i.e., neutrons with the energy $E< 10^{-7} $ eV
storaged in a trap. Previously similar experimental setup was used
in the search for   neutron-anti-neutron oscillations (see
\cite{7} and references therein).  The crucial  difference between
$n-n'$ and $n-\bar n$ oscillations is that $n'$ freely escapes from
the trap while $\bar n$ either annihilates on the trap walls or
gets reflected. Therefore formalism developed for the $n-\bar n$
oscillations cannot be  adjusted to treat $n-n'$ transitions.
Still the two processes have a common point. This is a problem of
a  correct quantum mechanical description of the UCN wave function
 (w.f.). Most often it is
assumed that the w.f. of the bottled UCN corresponds to a
stationary state of a particle inside a potential well \cite{8,9}.
Alternatively, other authors \cite{10} describe oscillations of
the trapped neutrons in the basis of the free plane waves. Both
pictures do not correspond to the physics of real experiments. The
process proceeds in time in three stages.

At the first stage the filling of the trap takes place, then the
neutrons are kept inside the  trap during the storage  time
(hundreds of seconds), and finally neutrons leave the trap to the
detectors. Therefore the w.f. undergoes  a complicated evolution
which hardly can be  described without resorting to
approximations. We shall first evaluate the neutron-mirror-neutron
oscillations using a stationary  w.f.  as the initial state w.f.
Then we shall do the same using  wave packet instead of a
stationary w.f.

The paper is organized as follows. We start in Section 2 with the
analysis of the oscillations in the stationary w.f. approach.
Transitions take place from one of the trap eigenstates. In
Section 3 transitions are considered in presence of a superimposed
magnetic field. A  general equation for the transition rate is
derived  and the limits of weak and strong field are considered.
Section 4 is devoted to the  wave packet formalism. The evolution
of the UCN wave packet (w.p.) is encoded using the trap Green's
function. Neutron-mirror-neutron transition rate is calculated. In
Section 5 the main conclusions are presented and open problems are
formulated. Appendix contains comparison between the infinite and
finite well models.

\section{Stationary Wave Function  Approach}

The problem of neutron-mirror-neutron  oscillations in free space
can be solved by diagonalization of the time-dependent two-channel
Schrodinger equation with the result \cite{11} \be |\psi_{n'}
(t)|^2 =\frac{4\varepsilon^2}{\omega^2+4\varepsilon^2}\exp
(-\Gamma_\beta t) \sin^2 \left(\frac12\sqrt{\omega^2+
4\varepsilon^2}t\right),\label{1}\ee where
$\omega=E_n-E_{n'}=|\mu_n|B$ is the energy difference between
neutron and mirror neutron due to superimposed magnetic field
(mirror neutron does not feel "our" magnetic field),
$\varepsilon=\tau^{-1}_{osc}$ is the mixing parameter,
$\Gamma_\beta$ is the neutron $\beta$-decay width. In arriving at
(\ref{1}) the spatial part of the w.f. was factored out making use
of the fact that in free space the w.f.-s of $n$ and $n'$ are of
the same form. In the trap, however, the situation is different:
the neutron is confined while for the mirror neutron the trap
walls do not exist. As already mentioned in the Introduction, the
description of the trapped UCN is a nontrivial problem. The naive
guess would be that inside the trap the neutron w.f. corresponds
to a discrete eigenstate.  Here we assume that the neutron w.f. is
that of a particle in a potential well with the boundary
conditions corresponding (in the first approximation) to a
complete reflection.

In order to make calculations tracktable and transparent we shall
consider the following simple model of a trap. Let it be a
one-dimensional square well of width $L=1m$ with walls at $x=0$
and $x=L$, i.e., the potential of the form \be U(x) =\left\{
\begin{array}{ll} V,&x<0\\0,& 0<x<L\\V,& x>L\end{array}\right.
.\label{2}\ee

The height of the potential well depends on the material with the
typical value $V=2\cdot 10^{-7}$ eV which will be used in what
follows. For such a well the limit for stored UCN velocity is $6.2
m/s$. The number of discrete levels in such a trap may be
estimated as \be M\simeq \frac{L\sqrt{2mV}}{\pi} \simeq
\frac{10^8}{\pi}.\label{3}\ee

We choose the  UCN energy to be   $E=0.8 \cdot 10^{-7}$ eV. This
energy  corresponds to a level with quantum number  $j\simeq
2\cdot 10^7$.  Positions and eigenfunctions of such highly excited
states in a finite-depth potential are very close to the same
quantities in the infinite well (except for the levels close to
the upper edge of the well; we not consider such levels). The
finite-depth corrections are considered in the Appendix. The
eigenvalues and eigenfunctions for the infinite well are \be
E_j=\frac{\pi^2 j^2}{2mL^2},~~ k_j=\frac{\pi j}{L},~~
j=1,2,3..\label{4}\ee \be \varphi_j(x) =\sqrt{\frac{2}{L}} \sin
k_j x.\label{5}\ee

Another important quantity characterizing highly excited states is
the classical frequency $\omega_{cl}$ \cite{10} \be \omega_{cl}
=\frac{\pi^2}{mL^2} j =\frac{2\pi}{\tau_{cl}}=\delta
E_j,\label{6a}\ee where $\tau_{cl}$ is the time of the classical
period and $$\delta E_j=E_{j+1} - E_j\simeq 0.8 \cdot
10^{-14}~{\rm eV}$$ is the level spacing. Levels with $j\gg 1$ are
almost equidistant. In the semiclasical limit we may also define
the trap crossing time $\tau$ \be \tau=\frac{\tau_{cl}}{2} =\frac{
mL}{k_j} \simeq 0.26 s\label{7a}\ee for $j=2\cdot 10^{7}$.
 Next we calculate
the rate of $(n-n')$ oscillations for the neutron at the $j-th$
discrete level. The neutron and mirror neutron w.f.-s in a
two-component basis  are \be \tilde \varphi_j(x)
=\sqrt{\frac{2}{L}} \sin k_j x\left(
\begin{array}{l} 1\\0\end{array}\right)\equiv \varphi_j(x)\left(
\begin{array}{l} 1\\0\end{array}\right)  ,\label{6}\ee \be \tilde f_p (x) =
\frac{1}{\sqrt{2\pi}} e^{ ipx}\left( \begin{array}{l}
0\\1\end{array}\right)\equiv  f_p (x) \left(
\begin{array}{l} 0\\1\end{array}\right),\label{7}\ee
where $-\infty <p <+\infty.$  The ($n-n')$ system is described by
the Hamiltonian \be \hat H=\hat H_0 + \hat W =\left(
\begin{array}{cc}
\frac{k^2}{2m}+U&0\\0&\frac{p^2}{2m}\end{array}\right) + \left(
\begin{array}{ll}0&\varepsilon\\
\varepsilon&0\end{array}\right).\label{8}\ee

The states (\ref{6}) and (\ref{7}) are the eigenstates of $\hat
H_0$, therefore it is convenient to use the interaction
representation. The probability to  find at time $t$ a mirror
neutron instead of a neutron reads

\be P_{nn'} =\int dp |\lan \tilde f_p|\exp \left \{ -i \int^t_0
dt' \hat W_{int} (t')\right\} |\tilde
\varphi_j\ran|^2,\label{9}\ee where $\hat W_{int} (t) = e^{i\hat
H_0t}\hat W e^{-i\hat H_0 t}$. In the first order of perturbation
theory we get
$$ P_{nn'} =\int dp |\lan\tilde  f_p| \int^t_0 dt' \hat W_{int}
(t') | \tilde \varphi_j\ran |^2=$$ \be =\varepsilon^2 \int dp |
\int^t_0 dt' e^{-i(E_j-E_p)t'}|^2 |\lan f_p|\varphi_j\ran
|^2,\label{10}\ee where $E_j=\frac{k_j^2}{2m},~~
E_p=\frac{p^2}{2m}.$
 The time-dependent integral is a standard
one \be w(E_p) =|\int^t_0 dt' e^{-i(E_j-E_p)t'}|^2 =\frac{4\sin^2
\left[\frac{(E_p-E_j)t}{2}\right]}{(E_p-E_j)^2}.\label{11}\ee The
overlap of the w.f.-s reads \be g_j(p) = |\lan
f_p|\varphi_j\ran|^2 = \frac{4k^2_j}{\pi L(p^2-k^2_j)^2} \sin^2
\left(\frac{pL+\pi j}{2}\right), ~~ j=1,2,...\label{12}\ee From
(\ref{10}), (\ref{11}) and (\ref{12}) we obtain \be P_{nn'}
=\varepsilon^2 \int^{+\infty}_{-\infty} dp g_j(p)
w(E_p).\label{13}\ee It is convenient to change integration from
$dp$ to $dE_p$ taking into account that $g(p) = g(-p)$.

Then \be P_{nn'} =2m\varepsilon^2 \int dE_p
\frac{g(E_p)w(E_p)}{p},\label{14}\ee where the factor 2 comes from
the fact that two plane waves $e^{\pm ipx}$ correspond to the same
energy $E_p$. Both functions $g(E_p)$ and $w(E_p)$ are peaked at $
E_p=E_j$. According to (\ref{12}) and (\ref{11}) the widths
$\Delta E^g_p$ and $\Delta E_p^w$ of the corresponding maxima are
\be \Delta E^g_p \simeq \pi/\tau,~~ \Delta E^w_p \simeq
4\pi/t,\label{15}\ee with $\tau$ being the trap crossing time. At
times $t\gg \tau$ we may substitute $g(E_p)/p$ by its value at
$p=k_j$ and take it out of the integral (\ref{14}). From
(\ref{12}) one gets $g(E_j) =L/4\pi$. The remaining integration in
(\ref{14}) can be extended to $(-\infty < E_p<+\infty)$ yielding
$2\pi t$. Collecting all pieces together we obtain \be P_{nn'}
=\varepsilon^2 \tau t.\label{16}\ee

At very short times  $t\ll \tau$ the function $w(E_p)$ becomes
smoother than $g(E_p)$. Hence $w(E_p)$ can be taken out of the
integral (\ref{14}). The remaining integral is time-independent
while $w(E_p) \sim t^2$. As a result $P_{nn'}\sim
\varepsilon^2t^2$ and we can not define the transition probability
per unit time \cite{11}. On the other hand, Eq.(\ref{16}) is valid
only for times shorter than the neutron $\beta $ -decay time
$t_\beta$ since we have define the eigenstate (\ref{6}) neglecting
the $\beta$-decay. The condition  $\tau \ll t \ll t_\beta$ was
with a fair accuracy  satisfied in experiments \cite{4,5}.

\section{Stationary approach with magnetic field included}

The search for $n-n'$ oscillations is experiments with bottled UCN
is based on the comparison of UCN storage with and without
superimposed magnetic field \cite{4,5}. It is assumed that there
is no mirror magnetic field in the laboratory and therefore the
interaction of the neutron with magnetic field lifts the
degeneracy  and thus suppresses the oscillations.

In magnetic field $B$ the energy of the trapped neutron becomes
equal to \be E_j =\frac{k^2_j}{2m} + \mu B,\label{17}\ee where
$\mu =-\mu_n = 1.91  \mu_N (\mu_N=e/2m_p).$

Inclusion of the magnetic field does not alter the functions
$w(E_p)$ and $g(p)$ given by Eq.-s (\ref{11}) and (\ref{12}).
There is, however, an important difference between our present
considerations and the previous section. As we see from (\ref{17})
$w(E_p)$ now peaks at $p=\pm \sqrt{k^2_j+ 2 m\mu B}$ while the
maximum of $g(p)$ is as before at $p=\pm k_j$. As  a result
instead of (\ref{14}) we obtain \be P_{nn'}
=\frac{4\varepsilon^2t}{(\mu B)^2 \tau \sqrt{1+\frac{2m\mu
B}{k^2_j}}}\left\{ \begin{array}{ll} \cos^2
\frac{k_jL}{2}\sqrt{1+\frac{2m\mu B}{k^2_j}},& j=1,3,...\\
\sin^2 \frac{k_jL}{2}\sqrt{1+\frac{2m\mu B}{k^2_j}},&
j=2,4...\end{array}\right.\label{18}\ee

This equation can be simplified taking into account that the
quantities $(\mu B)$  and $k^2_j/2m$ differ by many orders of
magnihides. In experiments \cite{4,5} the value of the  magnetic
field varied in the interval $(1-2)nT\leq B\leq ({\rm few})\mu T$
which corresponds to $10^{-16}$ eV $\la \mu B\la 10^{-13}$ eV,
while $k^2_j/2m\simeq 10^{-7}$ eV (note that the unshielded Earth
magnetic field corresponds to $\mu B\simeq 3\cdot 10^{-12}$ eV$
\ll \frac{k^2_j}{2m}$).

Therefore Eq. (\ref{18}) easily reduces to \be P_{nn'} \simeq 4
\varepsilon^2 \frac{t}{\tau} \frac{\sin^2(\frac12\mu B\tau)}{(\mu
B)^2},\label{19}\ee where $\tau$ is the trap crossing time. For
our model of the trap described in section 2 we have $
\tau/2\simeq 2\cdot 10^{14}$ eV$^{-1}$. Therefore in the limit of
weak magnetic field $B\simeq n T$ Eq. (\ref{19}) yields \be
P_{nn'} \simeq \varepsilon^2 \tau t,\label{20}\ee as expected (see
(\ref{16})). In the opposite  limit of strong magnetic field
$B\simeq $ (few) $\mu T$ we have to take into account that the
quantities $\tau$ and $B$ in (\ref{19}) experience fluctuations
leading to rapid oscillations of the function $\sin^2 (\frac12 \mu
B\tau)$. In particular, the crossing time $\tau$ may vary either
due to changes of $L$ at each cossision, or due to variations of
the neutron velocity. Substituting the rapidly oscillating
quantity in (\ref{19}) by its mean value equal to $1/2$ we obtain
the equation describing the neutron-mirror-neutron transitions in
strong magnetic field \be P_{nn'} =\varepsilon^2 \frac{2t}{(\mu
B)^2 \tau}.\label{21}\ee

\section{The wave packet approach}

We now turn to the question formulated in the Introduction, namely
to the problem of the UCN w.f. evolution and to the calculation of
the oscillations in the wave packet  approach. In order to get
physically transparent results and to avoid numerical calculations
suited to a concrete experiment we assume that UCN  coming to the
trap from the source are described by the Gaussian wave packets
(w.p.) \cite{12}.

The w.p. moving from the  left and for $t=0$ centered at $x=0$ is
given by the following expression

\be \Psi_k (x,t=0) =(\pi a^2)^{-1/4} \exp \left\{
-\frac{(x-x_0)^2}{2a^2} + ikx\right\},\label{22}\ee where $a$ is
the width of the w.p. and $k$ is its central momentum. The
normalization of the w.p. (\ref{21}) corresponds to one particle
in the entire one-dimensional space, \be \int^{+\infty}_{-\infty}
dx \left| \Psi_k(x, t=0)\right|^2=1.\label{23}\ee

Let the UCN energy be equal to the value chosen in Section 2,
$E=0.8\cdot 10^{-7}$ eV, and let the beam resolution be equal to
$\Delta E/E=10^{-3}$. Thus the set of parameters to be used
is\footnote{The problem of the choice of the w.p. parameters will
be addressed in the next Section.} \be E=0.8\cdot 10^{-7}~ {\rm
eV}, ~~ \lambda= \frac{2\pi}{k}\simeq 10^{-5}~{\rm cm},~~ a\simeq
3.2\cdot 10^{-3}~{\rm cm}.\label{24}\ee

The condition $a\gg \lambda$~ ensures the localization of the w.p.
We remind that the above  value of $E$ corresponds to the level
$E_j$ with a very high quantum number $j\simeq 2\cdot 10^7$.  Next
we estimate the number of levels within $\Delta E$. One has \be
\Delta j=\frac{\Delta E}{\omega_{cl}}=\frac{v(\Delta
k)}{\omega_{cl}} =\frac{L}{\pi a}\simeq 10^4.\label{26}\ee

The large number of levels forming the w.p. is a necessary
condition for the trapped w.p. to be localized (in free space this
condition reads $a\gg\lambda$, see above). The time evolution of
the initial w.p. (\ref{22}) proceeds according to the following
law \be \Psi_k(x,t) =\int dx' G(x,t; x',0) \Psi_k
(x',0),\label{27}\ee where $G(x,t; x',t')$ is the trap Green's
function. In the infinite well approximation we may use the
spectral decomposition of the Green's function over the set of
eigenfunctions (\ref{5}) and write \be \Psi_k (x,t)
=\sum^\infty_{j=1} e^{-iE_jt}\varphi_j (x) \int^L_0 dx'
\varphi^*_j(x') \Psi_k(x',0).\label{28}\ee The width of the w.p.
(\ref{28}) increases with time according to
$$a'=a\left[ 1+\left(\frac{t}{ma^2}\right)^2\right]^{1/2} \simeq
a\left(\frac{t}{ma^2}\right),$$ where for our model the spreading
time is $ma^2\simeq 1.7\cdot 10^{-2}s$ and $t/ma^2 \simeq 60
t(s)$. A so-called collapse time $t_c$ \cite{13} corresponds to
$a'=L$ and is equal to $t_c\simeq 500 s$. At $t=t_c $ the w.p.
spreads uniformly over the entire well and the stationary regime
considered in Section 2 sets in \cite{14}. We note in passing that
there is another time scale in the problem, the so-called revival
time $t_{rev} = 4 mL^2/\pi \simeq 2.10^7s$ when the w.p. regains
its initial shape --see \cite{13} and references therein.

The initial w.p $\Psi_k(x,0)$ contains only right running wave --
see (\ref{22}). The trapped w.p. (\ref{28}) contains both right
and left running waves, i.e., it correctly describes reflections
from the trap walls. We assume that the point $x_0$ (see
(\ref{22}) is not in the immediate vicinity of the  trap walls,
i.e., $x_0$ is at least few times of $a$ away from the walls. Then
the integration in (\ref{28}) may be  extended to the entire
one-dimensional space . This yields
$$ F(k,k_j; L,a,x_0) \equiv\int^{+\infty}_{-\infty}
dx'\varphi^*_j(x') \Psi_k(x',0)
 =i\left(\frac{a\sqrt{\pi}}{L}\right)^{1/2}\times $$
\be \times \left\{ \exp \left[ -\frac{a^2(k-k_j)^2}{2}+i(k-k_j)
x_0\right]-\exp \left[...k_j\to
-k_j...\right]\right\}.\label{29}\ee

Then we can calculate the transition probability $P_{nn'}$
following the procedure described in Section 2. Instead of the
w.f. (\ref{6}) we now have \be \Psi_k (x,t) =\sum^\infty_{j=1}
e^{-iE_jt} \varphi_j(x) F_j(k),\label{30}\ee with $F_j(k)$ being
the shorthand notation for the function $F_j(k,k_j; L, a, x_0)$
defined by (\ref{29}). The normalization condition for $F_j(k)$
reads \be \sum_j | F_j(k)|^2 =1.\label{31}\ee

In line with (\ref{13}) and following the arguments presented
after (\ref{14}) we write $$ P_{nn'} =\varepsilon^2 \sum_{j,l} F_j
(k) F^*_l(k) e^{\frac{i}{2}(E_l-E_j)t}\times
$$
\be \times \int^{+\infty}_{-\infty} dp \left[ \frac{2\sin
\frac{(E_p-E_j)t}{2}}{(E_p-E_j)}\right]\left[ \frac{2\sin
\frac{(E_p-E_l)t}{2}}{(E_p-E_l)}\right]\lan f_p|\varphi_j\ran \lan
\varphi_l|f_p\ran.\label{32}\ee

Consider first the contribution  $P^{(1)}_{nn'}$ of the  diagonal
terms with $j=l$. We have
$$ P_{nn'}^{(1)} =\varepsilon \sum_j | F_j (k)|^2
\int^{+\infty}_{-\infty} dp \frac{ 4\sin
\frac{(E_p-E_j)t}{2}}{(E_p-E_j)^2}\lan f_p|\varphi_j\ran \lan
\varphi_j|f_p\ran=$$ \be = \varepsilon^2\sum_j | F_j (k)|^2
\frac{2m}{k_j} 2\pi t \frac{L}{4\pi} =\varepsilon^2 \lan \tau \ran
t,\label{33}\ee with $\lan \tau\ran$ being the weighted crossing
time \be \lan \tau \ran = \sum_j |F_j(k)|^2 \tau
(k_j),\label{34}\ee and $\tau(k_j)= mL/k_j$. Next we turn to the
contribution $P^{(2)}_{nn"}$ of the nondiagonal terms in ({32}).
In this case we are dealing with a two-hump function with maxima
at $E_p=E_j$ and $E_p= E_l$. Therefore we may write

\beq  P^{(2)}_{nn'} &=&4 \pi m\varepsilon^2 \left\{ \sum_j
\frac{F_j(k)}{k_j} \sum_l F^*_l (k) e^{\frac12(E_l-E_j)t}\left[
\frac{2\sin \frac{(E_j-E_l)t}{2}}{(E_j-E_l)}\right] \right.\times
\nonumber \\
  &\times&
  \left. \lan f_j|\varphi_j\ran \lan\varphi_l|f_j\ran+(j\leftrightarrow l) \right \}.\label{32}\eeq
Replacing summation over $l$ by integration we obtain \be
P^{(2)}_{nn'} \simeq 8 \varepsilon^2 \sum_j |F_j|^2
\frac{m^2L^2}{k^2_j} =  8 \varepsilon^2 \lan
\tau^2\ran\label{36}\ee where \be \lan\tau^2\ran=\sum_j |F_j|^2
\tau^2_j.\label{37}\ee

Collecting the two contributions (\ref{33}) and (\ref{36})
together we get the final result \be P_{nn'} =\varepsilon^2 \lan
\tau\ran t \left( 1+ 8 \frac{\lan \tau^2\ran}{\lan \tau\ran
t}\right). \label{38}\ee

\section{Conclusions}

We have calculated the rate of neutron-mirror-neutron oscillations
 for trapped UCN. Two types of the UCN w.f.-s were used: the stationary solution
 for a particle inside a potential well and the Gaussian w.p. Calculations were
 performed in the first-order perturbation theory. This
 approximation is legitimate provided
$P_{nn'}\ll 1$. From (\ref{20}) and (\ref{21}) it follows that
this condition holds for $\tau_{nn'} \gg 7s$ and $\tau_{nn'}\gg
0.1s$ in the weak and strong magnetic fields correspondingly.
Obviously the  first order perturbation theory describes the
transition of the neutron into mirror neutron. The inverse process
appears only in the second order in $\varepsilon$. For the
analysis of the experiments \cite{4,5} first order perturbation
theory is a fair approximation.

The above analysis has been performed for a simple one-dimensional
trap. We think that such a model correctly describes the principal
features of the process. Generalization to the three-dimensional
rectangular trap is trivial. The simplest way to generalize our
result to the trap with arbitrary geometry is to substitute the
crossing time $\tau$ by the effective crossing time corresponding
to a given trap geometry.

Experimental data \cite{4,5} were analyzed using the free space
equation (\ref{1}) with the time $t$ being limited by the crossing
time $\tau$.

Equation (\ref{1}) contains only time dependence since the spatial
parts of  $n$ and $n'$ w.f.-s were factored out using the fact
that in free space the coordinate  w.f.-s of $n$ and $n'$ have the
same form. For bottled UCN the situation is different. Neutron is
confined inside the trap while mirror neutron freely crosses  the
trap  walls. Therefore the use of the (\ref{1}) to describe
oscillations of  trapped UCN seems questionable\footnote{To
describe the free space experiments by Eq.(\ref{1}) one still has
to supplement it by a boundary condition at $x=0$ where the
reactor is placed. Otherwise at $t=\pi\tau_{nn'}/2$ the reactor
becomes a source of mirror neutrons. We are grateful to L.B.Okun
and M.I.Vysotsky for drawing our attention to this.}. However our
accurate approach justifies the analysis of the experimental data
based on Eq.(\ref{1}) \cite{4,5}. This can be explained by the
semiclassical character of the UCN motion inside the trap of
macroscopic size. In the stationary approach the typical UCN
energy corresponds to the states with $j\gg1$, i.e., to the
semiclassical part of the spectrum. In the w.p. formalism the
classical limit corresponds to $(\Delta j )\sim (j)^{1/2} \to
\infty$ \cite{16}. The w.p. (\ref{22}),(\ref{24}) is close to this
limit. For UCN with extremely low energy, $E\la10^{-16}$ eV, the
oscillation pattern changes. We shall consider this question
elsewhere. The fraction of UCN with  such energies in experiments
is negligible. Another point which deserves a dedicated study is
decoherence of the UCN w.f. and subsequent randomization of the
oscillation process. This might occur due to dephasing of the w.f.
caused by collisions with the trap walls and with the residual
gas. 

We would like to thank A.P.Serebrov who  drew our attention
to the problem and with whom B.K. had numerous enlightening
discussions. Useful remarks were gained from L.B.Okun,
I.B.Khriplovich, A.Gal, O.M.Zherebtsov, A.I.Frank, V.A.Novikov.
One of us (B.K.) thanks Yuri Kamyshkov for the  hospitality and
support extended at the International Workshop on B-L Violation at
Berkeley in September 2007. Also B.K. thanks V.A.Gordeev and the
organizers of the XLII St.-Petersburg Winter School were
preliminary presentation of this work was made. Financial support
from  grants RFBR 06-02-17012, NS-4961.2008.2, NS-4568.2008.2,
RFBR 07-02-00830 and RFBR-08-02-00494 is gratefully acknowledged.
O.L. is also grateful to the Dynasty Foundation for the financial
support.


\section{Appendix}

 \setcounter{equation}{0} \def\theequation{A.\arabic{equation}}

 Calculations presented above were performed for the infinite well
 model of a trap. Here we consider the finite potential and show
 that there is only a minor difference between the two models.
 Consider the potential well defined by Eq. (\ref{2}). Matching the
 logarithmic derivatives of the w.f.-s at $x=0$ and $x=L$  we
 obtain the eigenvalue equation
 \be k'_jL=\pi j - 2 \arcsin
 \frac{k'_j}{\sqrt{2mV}},\label{A.1}\ee
 (the notation $k_j$ is kept  for $k_j=\pi j/L$). The small
 parameter in the problem is
 \be
 \delta = \left(\frac{2}{mVL^2}\right)^{1/2} \simeq 2\cdot
 10^{-8}.\label{A.2}\ee

 Expanding (\ref{A.1}) with respect to $\delta$ we obtain
 \be
 k'_j\simeq \frac{\pi j}{L} (1-\delta),~~ E'_j \simeq \frac{\pi^2
 j^2}{2 mL^2} (1-2\delta).\label{A.3}\ee
 Therefore the levels in the finite well are shifted relative to
 the infinite well levels by
 \be
 E_j-E'_j\simeq 4\cdot 10^{-15} {\rm ~eV}.\label{A.4}\ee
 From (\ref{A.3}) it follows that the spectrum in the finite well
 (\ref{2}) is the same as in somewhat wider infinite well
 \be
 L'=L(1+\delta).\label{A.5}\ee

 In the finite well the  w.f. penetrates into classically
 forbidden regions inside the trap walls. However
 neutron-mirror-neutron transitions inside the walls may be
 neglected since both the penetration depth $d$ and the collision
 time $\tau_{coll}$ are small: $ d\sim 10^{-6}$ cm, $\tau_{coll}
 \sim 10^{-8} s$.

\end{document}